\documentstyle[epsf,graphicx]{mn}

\def\g{$\Gamma$}
\def\x2{$\chi^{2}$}

\def\lunits{$\rm erg~s^{-1}~$}
\def\funits{$\rm erg~cm^{-2}~s^{-1}~$}

\newbox\grsign \setbox\grsign=\hbox{$>$} \newdimen\grdimen \grdimen=\ht\grsign
\newbox\simlessbox \newbox\simgreatbox \newbox\simpropbox
\setbox\simgreatbox=\hbox{\raise.5ex\hbox{$>$}\llap
     {\lower.5ex\hbox{$\sim$}}}\ht1=\grdimen\dp1=0pt
\setbox\simlessbox=\hbox{\raise.5ex\hbox{$<$}\llap
     {\lower.5ex\hbox{$\sim$}}}\ht2=\grdimen\dp2=0pt
\setbox\simpropbox=\hbox{\raise.5ex\hbox{$\propto$}\llap
     {\lower.5ex\hbox{$\sim$}}}\ht2=\grdimen\dp2=0pt

\begin{document}

\title[X-ray absorption in optically selected QSOs]
{The XMM-Newton/2dF survey VII. is there any X-ray absorption in
optically selected QSOs? }

\author[Akylas, Georgakakis \& Georgantopoulos]
{\Large A. Akylas$^{1,2}$,  A. Georgakakis$^1$, 
I. Georgantopoulos$^{1}$\\ 
$^1$ Institute of Astronomy \& Astrophysics, National Observatory of Athens, I. Metaxa
 B. Pavlou, Penteli, 15236, Athens, Greece \\
$^2$ Physics Department University of Athens, Panepistimiopolis, 
 Zografos, 15783, Athens, Greece \\  
}

\maketitle

\begin{abstract}

We explore the X-ray properties of optically selected QSOs
spectroscopically identified in the course of the  2dF QSO survey 
(2QZ). 
Our main goal is to expand to higher redshifts previous findings 
suggesting the presence of a fraction  
of X-ray obscured sources among 
the low redshift optically selected broad line 
AGN population.
The X-ray data are from the wide field ($\sim$2.5 deg$^2$)
shallow $[f( 0.5 - 8\,\rm keV) \approx \times 10^{-14}$ \funits]
XMM-{\it Newton}/2dF survey. A total of 96 2QZ QSOs overlap with the  
area covered by our X-ray survey. 66 of them have X-ray counterparts
while 30 remain undetected  in our X-ray survey.
The 66 X-ray
detected QSOs have a mean photon index of $\approx2$ suggesting little
or no X-ray obscuration for most of these sources. Individual X-ray
spectral fittings reveal only 1 source (intrinsic $\rm L_{X}(0.5-8~ keV)
\rm \sim
10^{44}~ erg ~ s^{-1}$ at $z=0.82$) that is likely to be obscured
($\rm N_H \approx 10^{23} \, cm^{-2}$) at the 90\% confidence
level. Additionally, there are 9 2QZ sources that show evidence 
for moderate absorption (mean observed $\rm N_H$ of
$\rm \approx10^{21}\,cm^{-2}$). For the 30 QSOs that remain undetected in
our X-ray  survey we use stacking analysis to estimate a mean hardness
ratio of $-0.59\pm0.11$ suggesting that the bulk of this population
has $\rm N_H$ consistent with the Galactic value. However, we cannot
exclude the possibility that some of these sources have enhanced
photoelectric absorption that is not revealed in the mean stacked
spectrum. We estimate a lower limit to the fraction of optically
selected  QSO with X-ray absorption of about 10\% (10 out of 96).  
\end{abstract}

\begin{keywords}

Galaxies: active -- quasars: general -- X-rays: general

\end{keywords}

\newpage

\section{INTRODUCTION}

Deep X-ray surveys in the 2-10\,keV band carried out with the {\it
Chandra} and  the XMM-{\it Newton} have resolved up to 90 per cent of
the hard (2-10\,keV) X-ray Background (XRB) into discrete sources to a
flux limit of  about $2\times10^{-16}$ \funits\ (Mushotzky et
al. 2000; Brandt et al. 2001; Giacconni et al. 2002). Follow-up
observations have shown that the X-ray sources in  these surveys are a
heterogeneous mix comprising broad line QSOs, Seyfert 1 and 2 type
systems, passive galaxies and optically faint sources ($I>24$\,mag)
whose nature remains elusive (Alexander et al. 2001; Barger et
al. 2001; Fiore et al. 2004; Georgantopoulos et al. 2004). These
observations are in stark contrast to the X-ray background (XRB)
population synthesis models, predicting a single dominant population
of heavily obscured AGNs (Comastri et al. 1995; Fiore et
al. 2004). Indeed, only a few cases of obscured (type-2; Norman et
al. 2002; Stern  et al. 2002) QSOs have been reported in the
literature to date, suggesting revision of the population synthesis
models.   

To this end, there is growing evidence for a population of broad-line
QSOs with little optical extinction (hence the broad optical lines)
but significant X-ray absorption (Akiyama et al. 2000; Brandt,  Laor
\& Wills al. 2000; Risaliti et al. 2001; Wilkes et al. 2002;  Akylas
et al. 2003). Although the apparent conflict between the optical and
X-ray absorption is still poorly understood, these objects may be an
important component of the X-ray background. Brandt et al.
(2000) argue that about 10\% of optically selected QSOs are X-ray weak
at soft energies with X-ray--to--optical flux density ratios 
$\alpha_{ox}>2$. Using the {\sc C\,iv} UV absorption spectral feature 
they argue that the X-ray weakness in these systems is most likely due
to absorption. Indeed, Gallagher et al. (2001) studied the X-ray
spectral properties of the soft X-ray weak population using {\it
ASCA} and {\it Chandra} data and found direct evidence for
significant obscuring column  densities in excess of $\rm 10^{22} \,
cm^{-2}$. On the basis of the studies above  Brandt et al. (2000)
suggest that selecting soft X-ray  weak galaxies ($\alpha_{ox}>2$)
produces samples with a high incidence of X-ray obscured
sources. However, broad line AGNs with significant X-ray absorption may
be present, albeit in smaller numbers, among the $\alpha_{ox}<2$
population.   

In this paper we address this issue by combining a homogeneous
sample of optically selected QSOs with a wide area ($\rm 2.5 \,deg^2$)
shallow ($f_X (\rm 0.5 - 8\, keV) \approx 10^{-14}\, erg~ \, s^{-1}\,
cm^{-2}$) XMM-{\it Newton} survey (Georgakakis et al. 2003, 2004). The
QSO sample is compiled from the 2dF QSO survey (2QZ; Croom et
al. 2001) which is based on optical/UV colour selection of sources
with $b_J<20.85$\,mag. This is about 4.5\,mag fainter than the
magnitude limit of the Bright Quasar Survey (Schmidt \& Green 1983)
used by Brandt et al. (2000) to explore the properties of soft X-ray
weak AGNs. 
The main goal of our study is to  
expand to higher redshifts previous findings 
suggesting the presence of a fraction  
of X-ray obscured sources among 
the low redshift optically selected broad line 
AGN population,
irrespective of their $\alpha_{ox}$, using X-ray spectral
analysis. Such a population could play an important role in the XRB
synthesis models as well as in unified AGN schemes.   

In Section 2 we  describe the X-ray and the optical data used in the
present study. Section 3 outlines the X-ray data reduction while
in Section 4 we present the X-ray data analysis. The results are
discussed  in Section 5 and our conclusions are summarized  
in Section 6. Throughout this paper we adopt $\rm H_{o}=65\, km \,
s^{-1} \,Mpc^{-1}$, $\Omega_M=0.3$  and $\Omega_{\Lambda}=0.7$. 

\section{THE X-RAY SAMPLE}

The X-ray data are from the wide field (18 pointings) bright (2-10\,ks 
per pointing) XMM-{\it Newton}/2dF survey. The observations are
carried out near the North (9 fields) and the South (9 fields)
Galactic Pole regions. 
We have excluded from our analysis 1 northern and 4 
southern XMM-{\it Newton} fields
suffering from strong particle background (see
Georgakakis et al. 2003, 2004). 
The remaining 13 fields cover a total
area of $\, \sim2.5$ $\rm deg^2$ and overlap with the 
2QZ survey.
The 2QZ is a large-scale  
spectroscopic program designed to follow optically selected QSOs using
the 2dF multi-fibre spectrograph on the 4-m Anglo-Australian Telescope
(AAT). The complete 2dF catalogue covers a total area of 740\,deg$^2$
and comprises about 25\,000 QSOs in the magnitude range $\rm 18.25<\
b_{j}< 20.85$\,mag.  The 2QZ spectra cover the wavelength range
3700--7900~\AA. A total of 96 2QZ QSOs overlap with
these fields.

The X-ray sources are detected in the 0.5-8\,keV energy band using a
threshold of 5$\sigma$. We detect 521 X-ray sources in total.   
The cross correlation between the X-ray data and the 2QZ 
catalogue reveals 66 matches within a distance less than
5\,arcsec. This represents about 70 per cent of the 2QZ 
QSOs in our fields. These optically identified Broad Line QSOs 
are  located both in the North (33 sources)  and the South (33
sources) Galactic Pole regions  spanning the redshift range of
$0.5<z<3$. We also use radio data from the FIRST (Faint Images of the
Sky  at Twenty Centimeters; Becker et al. 1995) and the NVSS (NRAO VLA
Sky Survey; Condon et al. 1998) radio surveys to identify our X-ray
data with radio sources. The observations are carried out at 1.4 GHz
with a limiting  flux density of 1 and 2.5\,mJy for the FIRST and the
NVSS respectively. Within a 10\,arcsec distance there is no coincidence
of our optically selected QSOs with radio sources.

\begin{figure}
\centering
\rotatebox{0}{\includegraphics[height=7.0cm]{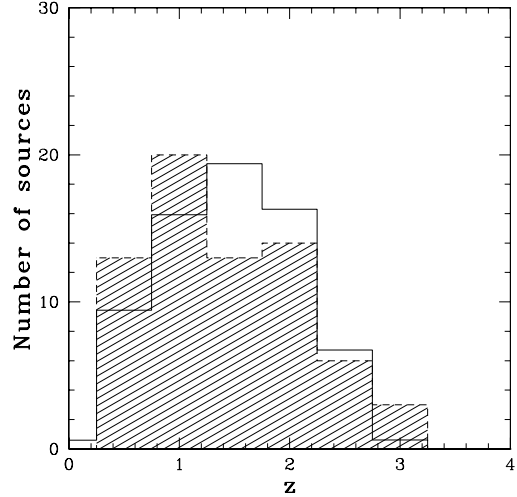}}   

\caption{The redshift distribution of 66 optically selected QSOs 
(shaded histogram) in comparison with the normalized redshift
distribution of the full 2QZ QSO catalogue (unshaded histogram).}
\label{redshift_dist}
\end{figure}

In Figure \ref{redshift_dist} we compare the redshift distribution of
the 66 QSOs of our sample (shaded histogram) with that of the
(normalized) full 2QZ sample. A Kolmogorov-Smirnoff test suggests that
the two datasets are drawn from the same parent population 
at the 90 per cent confidence level. The median redshift of the 2QZ
QSOs with and without X-ray counterparts is 1.3 and 1.7 respectively. 
The mean 2500\,\AA\ luminosity density of the two populations above is 
estimated  to be $2.2\times10^{30} \rm
\,erg~\, s^{-1}\, Hz^{-1}$ and $3.4\times 10^{30}\,\rm
erg~\, s^{-1}\, Hz^{-1}$ respectively (see section 5 for
details).  

\section{THE X-RAY DATA REDUCTION}

The X-ray data have been obtained with the EPIC (European Photon
Imaging Camera; Str\"{u}der et al. 2001 and Turner et al. 2001)
cameras on board the XMM-{\it Newton} operating in full frame mode
with the thin filter applied. The data have been analyzed using 
the Science Analysis Software ({\sc sas 5.3}). Event files for both
the PN and the MOS detectors   have been produced using the {\sc
epchain} and {\sc emchain} tasks of {\sc sas} respectively. The event
files were screened for high particle  background periods by rejecting
times with 0.5-10\,keV count rates higher than 25 and 15\,cts/s  for the  
PN and the two MOS cameras respectively.

In our analysis we have dealt only with events corresponding to
patterns 0-4 for the PN and 0-12 for the MOS instruments. To increase
the signal-to-noise ratio and to reach fainter fluxes we have merged
the PN and the MOS event files into a single event list using the {\sc
merge} task of {\sc sas}. We have extracted images and background
maps in three different energy bands  0.5-8 (total), 0.5-2 (soft),
and 2-8\,keV (hard)  for both the individual and the combined event
files. Exposure maps and bad pixel masks have also been constructed in
the above energy bands to take into account the vignetting  effect and
the presence of hot CCD pixels and gaps between the CCDs. We use the 
more sensitive (higher S/N ratio) merged image for source  extraction
and flux estimation, while the individual PN and MOS images are used
to calculate hardness ratios. This is because the interpretation of
hardness ratios is simplified if the extracted count rates  are from
one detector only. A small fraction of sources lie close to masked
regions (CCD gaps or hot pixels) on either the MOS or the PN
detectors. This may introduce errors in the estimated source
counts. To avoid this bias, the source count rates (and hence the
hardness ratios and the flux) are estimated  using the detector (MOS
or PN) with no masked pixels in the vicinity of the source.

The source counts for all the images are estimated within an 18\,arcsec
circle area. This area includes at least 70 per cent of the X-ray
source photons at off-axis angles less than 10\,arcmin. The source
counts are divided with the appropriate  exposure time to correct for
the vignetting effect. For the encircled energy correction,
accounting for the energy fraction outside the aperture within which 
source counts are accumulated, we adopt the calibration
 given by the XMM-{\it Newton} Calibration
Documentation\footnote{http://xmm.vilspa.esa.es/external/xmm\_sw\_cal/calib 
\\ /documentation.shtml\#XRT}. We convert count rates to flux assuming an
absorbed  power law spectrum with  $\Gamma=1.7$ and  Galactic column
density  $\rm N_H=2\times 10^{20}~cm^{-2}$ appropriate for these
fields (Dickey \& Lockman 1990). 
The Energy Conversion Factor (ECF) is
obtained using {\sc pimms} software v3.3a. 
Adopting \g=2 will lower our 0.5-8 keV flux estimates by 20 per cent.
As discussed in section 5 this has a negligible effect in our  
analysis.
In the case of the
simultaneous detection in the mosaic image, the mean ECF is estimated
by weighting the ECFs of the individual detectors using the respective
exposure time. 

For the X-ray spectral analysis we produce the individual spectra
files using the {\sc sas} task {\sc evselect}. The background spectra
files are extracted from every image independently, using regions free
from sources. The response matrices and the auxiliary files  are
produced using the {\sc sas} tasks {\sc rmfgen} and {\sc arfgen}   
respectively. We have used the {\sc xspec} v11.2 software to fit the
data.  All the quoted errors correspond to the 90 per cent confidence 
level.

\section{DATA ANALYSIS} 

\subsection{Hardness ratios}
\begin{figure}
\centering
\rotatebox{0}{\includegraphics[height=7.0cm]{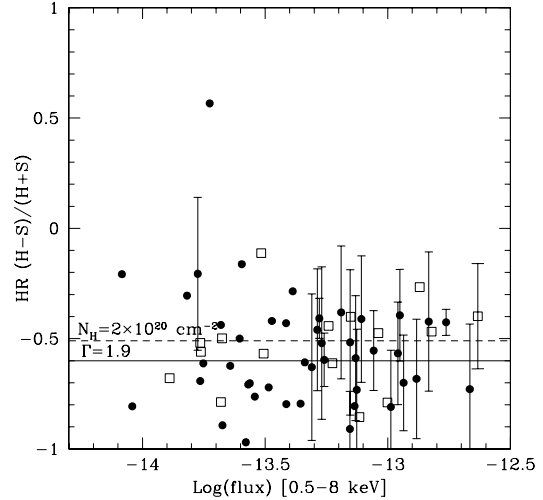}}   
\caption{
Hardness radio against 0.5-8\,keV flux. Filled circles denote the HR
values obtained using the PN data, while open boxes are for HR values  
obtained using the MOS data. The solid and dashed lines show,
respectively, the expected PN and MOS HR values for a power-law model
with  $\Gamma=1.9$ and $\rm N_H=2\times 10^{20}~ \rm cm^{-2}$. 
The error bars correspond to the 90 per cent confidence level.
For clarity, points with uncertainties greater than 0.35 are plotted
without error bars.
}
\label{hr_f}
\end{figure}

In Figure \ref{hr_f} we plot the hardness ratio (HR) 
as a function of the 0.5-8\,keV flux. 
By definition (HR=H-S/H+S, where H and S are 
the net counts in the 2-8 keV and 0.5-2 band respectively)
less negative HR values suggest less soft energy ($<$2 keV) 
photons compared to the harder energies ($>$2 keV) 
in the X-ray spectra. 
This can be attributed to the presence of a larger column 
density.
A total of eight sources have zero or negative net
counts in  the hard band. For these sources we set the HR to
--1. These sources are not plotted in Figure \ref{hr_f}. Due to the
presence of bad pixels,  the source counts are not always estimated
from  the same instrument (see section 3). In order to distinguish 
between the two different cases we use filled circles (42 points) 
and open boxes (16 points) to plot the HR values obtained using the PN
and MOS event files respectively. For clarity we plot the 90 per cent
error bars only when  they are smaller than 0.35. The two horizontal
lines show the expected HR for a power-law model with  $\Gamma=1.9$
and $\rm N_H=2\times 10^{20}~ \rm cm^{-2}$ for the PN (solid line) and
the MOS (dashed line). Note that the difference between the two lines
vanishes as we  move toward positive HR values (i.e. higher column
densities  $>5\times 10^{21}~ \rm cm^{-2}$).

\begin{figure}
\centering
\rotatebox{0}{\includegraphics[height=7.0cm]{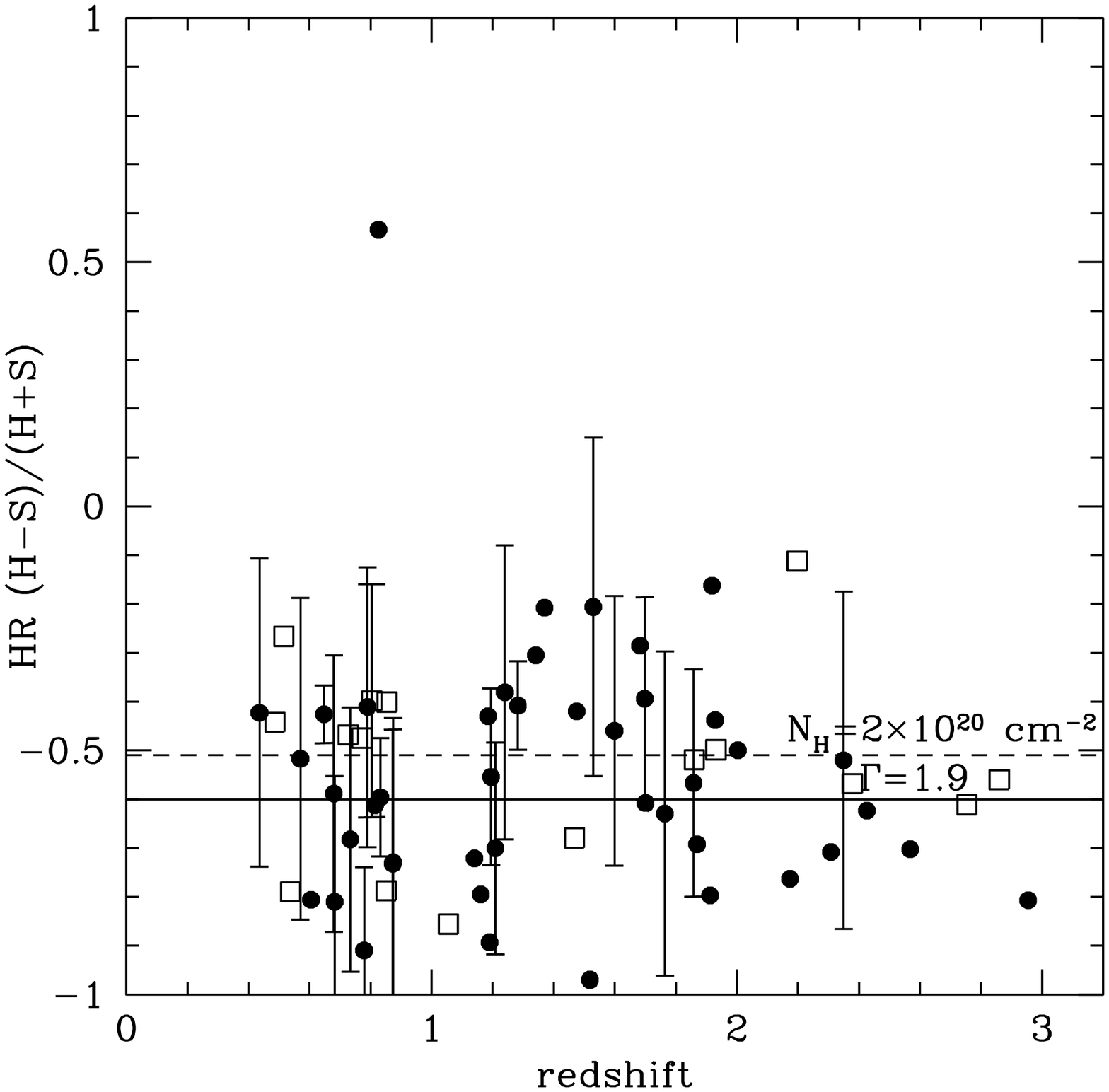}}   
\caption{Hardness radio against redshift. Filled circles denote the HR
values obtained  using the PN data, while open boxes are for HR values  
obtained using the MOS data. The solid and dashed lines show
respectively the expected PN and MOS HR values for a power low model
with  $\Gamma=1.9$ and $\rm N_H=2\times 10^{20}~ \rm cm^{-2}$. 
The error bars correspond to the 90 per cent confidence level.
For clarity, points with uncertainties greater than 0.35 are plotted
without error bars.
}
\label{hr_z}
\end{figure}

In Figure \ref{hr_f} there is no strong evidence of obscuration. 
Within the 90 per cent confidence level most of the HR values are 
consistent with a photon index of $\sim$ 1.9. There is only one source  
(\#38) which shows a very flat spectrum on the basis of the HR
value, albeit with large  error bar (HR=$0.57 \pm 0.60$). The
unweighed mean HR values for the PN and MOS data are  $\rm <HR_{\rm
PN}>=-0.55$ and $\rm <HR_{\rm MOS}>=-0.53$ corresponding to an
observed  column density of  $\rm N_H=5\times 10^{20}\, cm^{-2}$ for
$\Gamma=1.9$. The {\it observed} column density however, 
is lower than the {\it rest-frame} one because the $k$-correction
shifts the  absorption turnover to lower energies. The relation
between the intrinsic rest-frame and the observed column density
scales approximately as $(1+z)^{2.65}$ (Barger et al. 2002). Using
this relation the above mean hardness ratios correspond to a maximum
intrinsic column density of  $\rm N_H=2\times 10^{21} \, cm^{-2}$ at
$z=1$.  In Figure \ref{hr_z} we plot the HR as a function of 
redshift. There is no evidence for evolution of HR values with
redshift.

\subsection{Individual spectral fittings}

We attempt to further investigate the X-ray properties of the data
performing individual spectral fittings for all 66 sources. 
The poor count statistics do not allow the use of the standard
$\chi^2$ analysis. Instead we use the C-statistic technique (Cash
1979), which is proper for fitting spectra with limited number of
counts. Note that this method can be used to estimate
parameter values and confidence regions but does not provide a
goodness-of-fit (see Arnaud, George \& Tennant 1992). We fit
simultaneously the PN and the MOS data for each source 
in the energy range 0.2-8 keV. Due to the
presence of hot pixels and CCD gaps, especially in  the PN detector,
and because of a small  offset between the PN and the MOS field of
view, we are able to extract spectra from both the PN and the MOS
detectors  for 44 sources. There are 16 sources with MOS spectra
only and  6 sources with PN spectra only. 

We use an absorbed power law model to fit the data. First we fix
the photon index to 1.9 and allow both the $\rm N_H$ and the 
normalization to vary. The results are presented in Table 1.
There are only three sources (\#2, \#22, \#56) which appear to
have a column density above the Galactic one ($2\times 10^{20}\,\rm 
cm^{-2}$).  All the other $\rm N_H$ values are fully 
consistent with  
the Galactic value within the 90 per cent confidence level.
In this case the $\rm N_H$ values listed in Table 1 correspond to 
the 90 per cent upper limit. 
We also try to fit the data using an absorbed power law model 
with the $\rm N_H$ fixed to the Galactic value while the photon 
index and the normalization are free parameters. In this case a flat
photon index is an indication of obscuration. There is a
strong evidence for obscuration only in one case (source \#22).  
There are also nine sources (see Table 1) which may be obscured  
on the basis of their photon index value. For these sources  we find a
flat photon index,  $<$1.6. We note however, that within the the 90
per cent uncertainties the estimated $\Gamma$s are consistent with 
1.9.

\begin{table*} 
\caption{Spectral fitting results for 66 X-ray detected 2QZ  QSOs}
\begin{tabular}{ccccccccccccc}

\hline 

No & NAME & RA     & DEC    &  $ ^1f_{\chi}~(\times 10^{-14}$)  & z & $\rm ^2N_H~ (\times 10^{22}$) & $^3 \Gamma$ & $^4$DETECTOR \\
   &      &(J2000) &(J2000) &  \funits\  &   & $\rm cm^{-2}$               &	      & 	     \\
\hline

1 &   J005943.9-273831  & +00:59:44.0     & -27:38:33  &   4.59  & 1.70  &  $<0.048$  &	 $  2.07^{+0.43}_{-0.40}$& PN+MOS    \\
2 &   J005929.4-274339  & +00:59:29.3     & -27:43:42  &   8.76  & 1.19  &  $0.094^{+0.086}_{-0.049}$& $  1.42^{+0.28}_{-0.26}$& PN+MOS   \\
3 &   J005913.8-280652  & +00:59:13.9     & -28:06:54   &   1.77  & 0.81  &  $<0.150$	 &	 $  1.88^{+0.50}_{-0.46}$& PN+MOS    \\ 
4 &   J005903.5-275311  & +00:59:03.6     & -27:53:10   &   0.91  & 2.95  &  $<0.046$  &	 $  2.90^{+0.89}_{-0.74}$& PN        \\
5 &   J005900.7-273108  & +00:59:00.7     & -27:31:09   &   10.3  & 0.68  &  $<0.003$  &	 $  2.76^{+0.12}_{-0.20}$& PN+MOS    \\
6 &   J005859.1-273038  & +00:58:59.1     & -27:30:40  &   2.74  & 2.56  &  $<0.095$  &	 $  1.70^{+0.37}_{-0.40}$& PN+MOS    \\
7 &   J005858.5-280319  & +00:58:58.5     & -28:03:21   &   1.68  & 1.53  &  $<0.020$  &	 $  2.10^{+0.31}_{-0.40}$& PN+MOS    \\
8 &   J005855.0-273141  & +00:58:55.1     & -27:31:41  &   4.41  & 1.16  &  $<0.188$  &	 $  2.31^{+0.23}_{-0.33}$& PN+MOS    \\
9 &   J005850.8-280547  & +00:58:50.8     & -28:05:48   &   5.51  & 0.83  &  $<0.038$  &	 $  1.97^{+0.44}_{-0.25}$& PN        \\
10 &  J005843.7-273459  & +00:58:43.7     & -27:34:58  &   3.37  & 1.47  &  $<0.019$  &	 $  2.21^{+0.37}_{-0.34}$& PN+MOS    \\
11 &  J005836.1-273820  & +00:58:35.9     & -27:38:21  &   2.12  & 1.19  &  $<0.062$  &	 $  1.95^{+0.41}_{-0.56}$& PN+MOS    \\
12 &  J005831.5-273757  & +00:58:31.4     & -27:37:55  &   1.73  & 2.86  &  $<0.200$  &	 $  2.29^{+1.26}_{-0.93}$& MOS       \\
13 &  J005811.4-272636  & +00:58:11.8     & -27:26:31  &   2.11  & 2.48  &  $<0.016$  &	 $  1.05^{+1.14}_{-0.94}$& PN+MOS    \\
14 &  J005810.8-280817  & +00:58:10.7     & -28:08:21   &   2.09  & 1.92  &  $<0.120$	 &       $  2.34^{+1.49}_{-1.00}$& PN+MOS    \\
15 &  J005805.7-275005   & +00:58:05.8     &   -27:50:06   &   5.74  & 0.48  &  $<0.130$  &	 $  1.87^{+0.68}_{-0.62}$& MOS       \\
16 &  J005803.3-281211   & +00:58:03.5     &   -28:12:15   &   17.3  & 0.64  &  $<0.014$  &	 $  2.18^{+0.21}_{-0.28}$& PN        \\
17 &  J005747.5-275412   & +00:57:47.5     &   -27:54:10   &   5.26  & 1.28  &  $<0.020$  &	 $  1.95^{+0.57}_{-0.50}$& PN        \\
18 &  J005740.1-282311   & +00:57:40.0     &   -28:23:15   &  7.32  & 0.60  &  $<0.070$  &	 $  2.18^{+0.27}_{-0.25}$& PN+MOS    \\
19 &  J005734.9-272828   & +00:57:34.9     &   -27:28:29  &   2.87  & 2.17  &  $<0.100$ &	 $  2.01^{+0.57}_{-0.62}$& PN+MOS    \\
20 &  J005727.9-283107   & +00:57:27.9     &   -28:31:09   &   7.06  & 0.85  &  $<0.035$  &	 $  2.16^{+0.40}_{-0.37}$& MOS       \\
21 &  J005724.4-273201   & +00:57:24.4     &   -27:32:02   &   11.6  & 1.20  &  $<0.010$ &	 $  2.33^{+0.11}_{-0.22}$& PN+MOS    \\
22 &  J005701.1-272800   & +00:57:01.0     &   -27:28:02  &   1.88  & 0.82  &  $2.55^{+1.69}_{-1.38}$ & $  0.20^{+0.48}_{-0.77}$& PN+MOS \\ 
23 &  J005650.9-280955   & +00:56:50.9     &   -28: 9:56  &   9.15  & 0.77  &  $<0.095$ &	 $  1.75^{+0.47}_{-0.45}$& MOS       \\
24 &  J005648.9-282158   & +00:56:49.0     &   -28:22:01   &   2.70  & 2.30  &  $<0.065$	 &	 $  2.00^{+0.37}_{-0.34}$& PN+MOS    \\
25 &  J005634.8-281622   & +00:56:34.7     &   -28:16:22  &   2.11  & 1.93  &  $<0.048$ &	 $  2.67^{+1.00}_{-0.79}$& MOS       \\
26 &  J005612.4-275711   & +00:56:12.4     &   -27:57:12  &   15.1  & 0.72  &  $<0.030$ &	 $  2.03^{+0.20}_{-0.25}$& MOS       \\  
27 &  J005602.1-282658   & +00:56:02.2     &   -28:27:02   &   3.86  & 1.91  &  $<0.060$  &	 $  2.69^{+0.94}_{-0.80}$& PN+MOS    \\
28 &  J005549.3-280334   & +00:55:49.2     &   -28:03:39  &   2.55  & 1.01  &  $<0.033$ &	 $  2.40^{+0.32}_{-0.62}$& PN+MOS    \\
29 &  J005545.2-275734   & +00:55:45.1     &   -27:57:37  &   11.0  & 1.85  &  $<0.035$ &	 $  1.95^{+0.15}_{-0.17}$& PN+MOS    \\
30 &  J005543.2-280457   & +00:55:43.0     &   -28:04:58  &   7.48  & 0.87  &  $<0.048$ &	 $  1.99^{+0.23}_{-0.24}$& PN+MOS    \\
31 &  J005504.2-280732   & +00:55:04.1     &   -28:07:32 &   5.38  & 2.35  &  $<0.020$ &	 	 $  2.31^{+0.30}_{-0.34}$& PN+MOS    \\
32 &  J005459.1-281430   & +00:54:58.7     &   -28:14:31  &   7.01  & 0.77  &  $<0.057$ &	 $  2.21^{+0.56}_{-0.50}$& PN        \\
33 &  J005449.6-281125   & +00:54:49.2     &   -28:11:25  &   3.76  & 0.57  &  $<0.034$ &	 $  1.55^{+0.53}_{-0.65}$& PN+MOS    \\
34 &  J134512.3-003132   & +13:45:12.3     &   -00:31:31   &   4.10  & 1.68  &  $<0.070$  &	 $  1.89^{+0.37}_{-0.52}$& PN+MOS    \\
35 &  J134455.8-000116   & +13:44:55.8     &   -00:01:16   &   3.05  & 2.19  &  $<0.310$  &	 $  1.34^{+0.55}_{-0.83}$& MOS      \\
36 &  J134454.6-001908   & +13:44:54.9     &   -00:19:08   &   2.09  & 0.85  &  $<0.066$ &	 $  2.18^{+0.56}_{-0.50}$& MOS       \\
37 &  J134437.0+003054   & +13:44:37.1     &   +00:30:56    &   2.64  & 1.51  &  $<0.045$	&	 $  2.30^{+0.58}_{-0.43}$& PN+MOS   \\
38 &  J134427.9-003029   & +13:44:28.0     &   -00:30:32   &   0.83  & 1.37  &  $<1.600$  &	 $  1.40^{+1.45}_{-1.29}$& PN+MOS    \\
39 &  J134424.5-000617   & +13:44:24.8     &   -00:06:15   &   2.49  & 2.00  &  $<0.035$  &	 $  2.20^{+0.62}_{-0.67}$& PN+MOS   \\
40 &  J134420.8+000226   & +13:44:21.0     &   +00:02:29   &   1.72  & 1.87  &  $<0.055$  &	 $  2.08^{+0.57}_{-0.50}$& PN+MOS   \\
41 &  J134420.0-003111   & +13:44:20.1     &   -00:31:11   &   7.40  & 0.68  &  $<0.031$  &	 $  1.96^{+0.20}_{-0.27}$& PN+MOS    \\
42 &  J134414.0-002950   & +13:44:14.2     &   -00:29:52   &   9.97  & 0.53  &  $<0.004$ &	 $  2.89^{+0.27}_{-0.27}$& MOS       \\
43 &  J134353.4-000520   & +13:43:53.5     &   -00:05:18   &   3.27  & 1.14  &  $<0.290$  &	 $  1.52^{+0.43}_{-0.69}$& PN+MOS   \\
44 &  J134347.5-002336   & +13:43:47.6     &   -00:23:40   &   7.67  & 1.05  &  $<0.010$  &	 $  2.72^{+0.68}_{-0.58}$& MOS       \\
45 &  J134339.6+002937   & +13:43:40.0     &   +00:29:34    &   3.12  & 2.37  &  $<0.021$	&	 $  1.88^{+0.72}_{-0.74}$& MOS      \\
46 &  J134331.4+001108   & +13:43:31.4     &   +00:11:10   &   10.1  & 1.28  &  $<0.033$  &	 $  2.50^{+0.94}_{-0.72}$& PN+MOS   \\
  
\hline 	    
\end{tabular} 
\end{table*}

\begin{table*} 
\contcaption{}
\begin{tabular}{ccccccccccccc}

\hline 

No & NAME & RA     & DEC    &  $^1f_{\chi}~ (\times 10^{-14}$)  & z & $\rm ^2N_H~ (\times 10^{22}$) & $^3 \Gamma$ & $^4$DETECTOR \\
   &      &(J2000) &(J2000) &  \funits\  &   & $\rm cm^{-2}$               &	      & 	     \\
\hline
  
47 &  J134324.2-002030  & +13:43:24.2     &   -00:20:29   &   2.19  & 1.89  &  $<0.024$  &	 $  2.90^{+3.32}_{-1.20}$& PN        \\
48 &  J134323.6+001222  & +13:43:23.8    &    +00:12:21    &   21.6  & 0.87  &  $<0.031$	&	 $  2.06^{+0.14}_{-0.31}$& PN+MOS   \\
49 &  J134314.8+002528  & +13:43:14.9    &    +00:25:29    &   1.29  & 1.46  &  $<0.033$	&	 $  1.95^{+1.04}_{-0.77}$& MOS      \\
50 &  J134301.5-002951  & +13:43:01.6      &  -00:29:51    &   0.91  & 2.06  &  $<1.000$  &	 $  2.80^{+0.69}_{-1.50}$& PN+MOS    \\
51 &  J134256.5+000056  & +13:42:56.6     &   +00:00:57   &   23.3  & 0.80  &  $<0.048$  &	 $  1.81^{+0.15}_{-0.18}$& MOS      \\
52 &  J134255.4+000634  & +13:42:55.4     &   +00:06:36   &   14.7  & 0.43  &  $<0.052$  &	 $  2.07^{+0.73}_{-0.25}$& PN+MOS   \\
53 &  J134233.7-001148  & +13:42:33.7     &   -00:11:49   &   13.5  & 0.51  &  $<0.150$  &	 $  1.84^{+0.60}_{-0.54}$& MOS      \\
54 &  J134232.9-001551  & +13:42:33.0     &   -00:15:50   &   2.78  & 2.13  &  $<0.460$  &	 $  1.60^{+0.75}_{-0.67}$& PN+MOS    \\ 
55 &  J134219.1+000254  & +13:42:19.1     &   +00:02:56   &   3.85  & 1.18  &  $<0.150$  &	 $  1.86^{+0.95}_{-0.85}$& PN+MOS   \\
56 &  J134211.9+002950  & +13:42:12.0    &    +00:29:50    &   7.03  & 0.57  &  $0.087^{+0.083 }_{-0.062}$ & $  1.51^{+0.29}_{-0.29}$& PN+MOS  \\
57 &  J134156.8+003009  & +13:41:56.9    &    +00:30:10    &   6.46  & 1.24  &  $<0.046$	&	 $  1.94^{+0.25}_{-0.28}$& PN+MOS   \\
58 &  J134155.7-002233  & +13:41:55.9     &   -00:22:34   &   2.28  & 2.42  &  $<0.058$  &	 $  1.97^{+0.62}_{-0.58}$& PN+MOS    \\
59 &  J134142.8+001238  & +13:41:42.9    &    +00:12:39    &   7.81  & 0.79  &  $<0.100$	&	 $  1.80^{+0.30}_{-0.30}$& PN+MOS    \\
60 &  J134133.6-002704  & +13:41:33.8     &   -00:27:03   &   1.52  & 1.34  &  $<0.400$  &	 $  1.30^{+0.81}_{-0.69}$& PN+MOS    \\
61 &  J134127.9+003211  & +13:41:27.8    &    +00:32:13    &   4.90  & 1.76  &  $<0.070$	&	 $  1.95^{+0.20}_{-0.40}$& PN+MOS   \\
62 &  J134127.1+001413  & +13:41:27.2    &    +00:14:14    &   11.2  & 1.69  &  $<0.024$	&	 $  2.05^{+0.15}_{-0.23}$& PN+MOS    \\
63 &  J134122.8-002246  & +13:41:22.8     &   -00:22:43   &   2.54  & 1.91  &  $<0.060$  &	 $  1.98^{+0.41}_{-0.46}$& PN+MOS    \\
64 &  J134121.3-001353  & +13:41:21.6     &   -00:13:51   &   13.1  & 0.73  &  $<0.004$  &	 $  2.66^{+0.20}_{-0.25}$& PN+MOS    \\
65 &  J134059.1-001945  & +13:40:59.3     &   -00:19:45   &   1.72  & 1.86  &  $<0.140$  &	 $  2.01^{+1.09}_{-0.81}$& MOS       \\
66 &  J134041.4-001727  & +13:40:41.6     &   -00:17:24   &   5.95  & 2.75  &  $<0.095$  &	 $  1.87^{+0.50}_{-0.61}$& MOS       \\

\hline 	    

\end{tabular} 

\begin{tabular}{l}
$^1$ X-ray flux in the 0.5-8\,keV band assuming a power law model with $\Gamma$=1.7 
absorbed by a column density of $2\times 10^{20}~ \rm cm^{-2}$  \\
$^2$ These values are obtained using an absorbed power law model with $\Gamma$ fixed to 1.9. \\
$^3$ These values are obtained using an absorbed power law model with $\rm N_H$ fixed 
to $2\times 10^{20}~ \rm cm^{-2}$ \\ 
$^4$ The detector from which we have extracted the spectral files \\

\end{tabular}
\end{table*}     
	 
\begin{table} 
\caption{Average PN and MOS spectral fitting results 
for all the sources in our sample.}
\begin{tabular}{cccccccc}

\hline 

Sample & $\rm N_H (\times 10^{20})$  &  $\Gamma$  & $\chi^2$/dof \\
       & $ \rm cm^{-2}$ &	 \\
\hline 

$^1$PN 	  & $<1.1$  & $2.00^{+0.11}_{-0.06}$  & 146.7/115 \\
$^2$MOS	  & $<1.8$  & $2.07^{+0.10}_{-0.07}$  & 163.7/146 \\
PN+MOS &  $<1.1$    & $2.05^{+0.07}_{-0.06}$  & 313.87/263 \\
\hline 
  
$^1$ 51 sources \\
$^2$ 61 sources \\

\end{tabular} 

\end{table}

\subsection{Average spectra}
   
\begin{figure}
\centering
\rotatebox{-90}{\includegraphics[height=7.0cm]{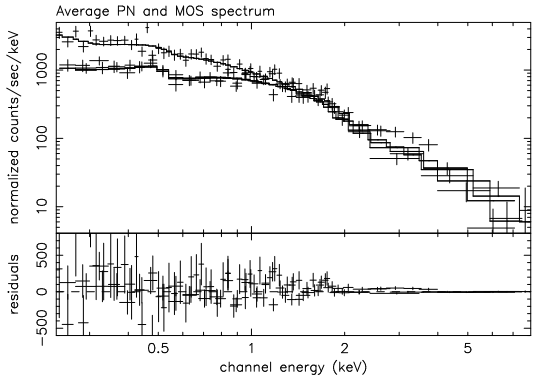}}   
\caption{
The average (0.2-8.0\,keV) spectrum, the best joint fit model and  the
residuals for the PN (top line) and the MOS (bottom lines) sources.
}
\label{average}
\end{figure}

Here we examine the average X-ray properties of our sample.
First we construct three  merged datasets adding respectively the 
individual PN (50), MOS1 (60) and MOS2 (60) spectral files, using the
ftool task {\sc mathpha}. These files are binned to give a minimum of
20 counts per bin so that Gaussian statistics can be applied. 
We  also use {\sc addarf} task in {\sc ftools} to create an  average
auxiliary file for each merged dataset. Since there are no differences
in the response matrices between the different CCDs we use one
individual response matrix for each merged dataset. Then we fit the PN
and the MOS data with an absorbed power law model. The merged spectral
files for the PN and the MOS detectors do not contain the same number
of sources so we first apply  the model to the PN and the MOS spectral
files separately. For comparison we also simultaneously fit the PN and
MOS data  ($\Gamma$ and $\rm N_H$ parameters for the PN and the MOS
are tied) allowing the normalization parameters for the PN and the MOS
to vary.   We present the results in Table 2.  Despite  the different
number of sources included in  each of these subsamples the results
are in good agreement. In Figure \ref{average} we plot the average
spectrum, the best joint fit model and the residuals for the PN (top
line), the MOS1 and the MOS2 (bottom lines)  spectral files.

X-ray background studies (e.g. Tozzi et al. 2001; Stern et al. 2002), 
have revealed a progressive hardening of 
the average photon index of the X-ray sources at lower fluxes.
This result has direct implication to the X-ray background 
synthesis models (Comastri et al. 1995).  
Here, we try to investigate this issue using our homogeneous 
selected sample of UVX QSOs.
We divide the data into two subsamples according 
to their flux. Sources with flux greater than $10^{-13}$ \funits\
constitute the bright sample and the remaining sources the faint sample.
We use the same model to simultaneously fit the PN and the MOS data. 
The best-fit parameters for the bright and the faint subsamples are 
$\Gamma=2.07^{+0.08}_{-0.06}$, 
$\rm N_H<0.7\times10^{20}~\rm cm^{-2}$ and $\Gamma=2.00^{+0.12}_{-0.09}$, 
$\rm N_H<2.1\times10^{20}~ \rm cm^{-2}$ respectively. 
At the fluxes probed here there is no
evidence for a change of the photon index. The above
results clearly show that on average the presence of obscuration in
our data is not important. However there are ten sources which present
a hard photon index ($<1.6$). The  average  photon index of these
sources is  $\Gamma=1.50^{+0.20}_{-0.20}$. Despite the large
uncertainties, this result may suggest the presence of a significant  
column density in these QSOs.
 
\begin{figure}
\centering
\rotatebox{0}{\includegraphics[height=7.0cm]{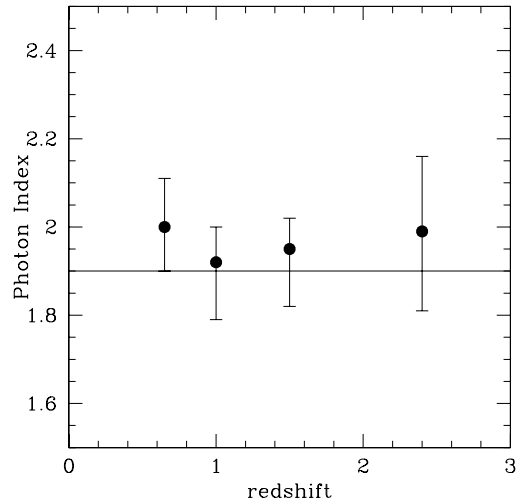}}   
\caption{
The average photon index versus mean redshift for the four redshift
bins. The estimated average values are consistent with 1.9 (solid
line).
} 
\label{gamma_evol}
\end{figure}

We have also divided the data into four subsamples based 
on the redshift. For this separation we adopt the following 
redshift intervals: z$<$0.8, 0.8$<$z$<$1.2, 1.2$<$z$<$1.8 
and z$>$1.8. 
These subsamples contain 16, 15, 16 and 22 sources respectively. 
We simultaneously fit the PN and the MOS data for each subsample 
applying a power law model. Figure \ref{gamma_evol} plots the average
$\Gamma$ values at the mean redshift of each bin. The solid line   
corresponds to $\Gamma=1.9$. 
There is no evidence for a trend of the spectral slope 
with redshift in agreement with Figure 3. All the average values 
are consistent with the mean slope of unobscured local AGNs
($\Gamma\sim1.9$, e.g. Nandra \& Pounds. 1994) 
at the 90 per cent confidence level.  
This result confirms previous findings (e.g. Vignali et al. 2003b,
Piconcelli et al. 2003) suggesting that the accretion
mechanism in radio quite QSOs is the same at any redshift.
We also note that the presence of a reflection component 
should significantly flatter the average spectrum of QSOs above z=2 
(see Akiyama et al. 2000). However the existence 
of this component at the luminosities probed here
($\overline{L_{X}}(\rm 0.5-8~keV)=8\times 10^{44}$ 
{\lunits} for $\bar{z}$=2.2) is not 
seen.

\begin{figure}
\centering   
\rotatebox{0}{\includegraphics[width=7.0cm]{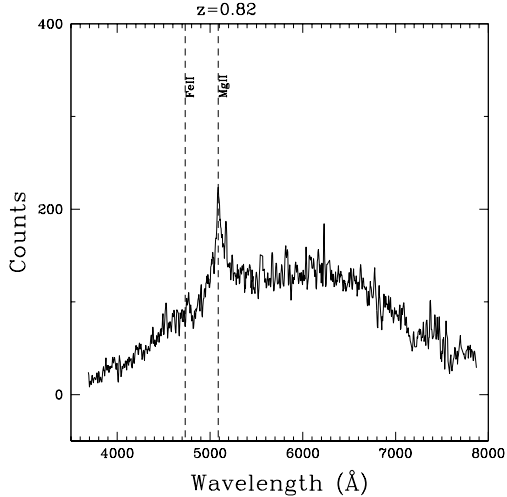}}   
\caption{The 
2DF optical  spectrum (not calibrated in flux) of the absorbed QSO ($\#22$)
}
\label{optical}
\end{figure}

\begin{figure}
\centering   
\rotatebox{0}{\includegraphics[width=7.0cm]{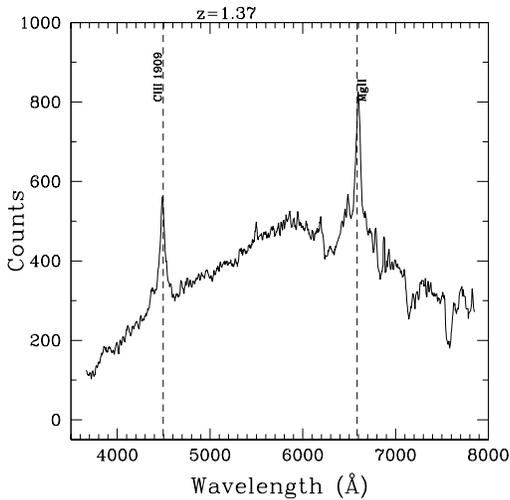}}   
\caption{The 2DF optical  spectrum (not calibrated in flux), of source $\#38$ 
which presents $\alpha_{ox}>2$}
\label{optical1}
\end{figure}

\begin{figure}
\centering
\rotatebox{0}{\includegraphics[height=7.0cm]{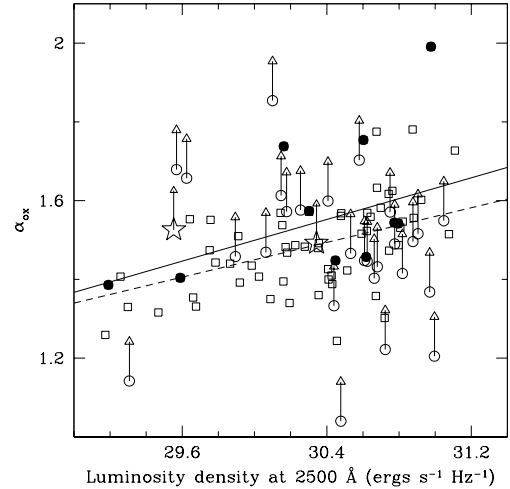}}   
\caption{
The two point optical/X-ray index, $\alpha_{ox}$, against 2500 \AA\
luminosity density for the 96 2QZ QSOs.  The 30 X-ray undetected QSOs
are shown as lower limits. The stars are BAL QSOs on the basis  
of the 2QZ classification scheme. The filled circles denote the 
X-ray detected QSOs with a flat ($<$1.6) best fit power law photon
index and the open squares shows the  66 X-ray detected QSOs. 
The solid line corresponds to our best linear fit to the data. The
dashed line is the best fit model presented by Vignali et al.
(2003a).
}
\label{aox} 
\end{figure}

\section{DISCUSSION}

We first attempt to identify obscured 2QZ QSOs using X-ray spectral
fitting analysis. Using the C-statistic and assuming a power-law model
with $\Gamma=1.9$ we find evidence for obscuration above the $90$ per
cent level in three sources (\#2, \#22 and \#56). Using a flatter  
photon index, $\Gamma=1.7$, only one source (\#22) remains  
significantly absorbed at the 90 per cent confidence level. The rest
frame column density of this object is estimated to be $\rm
N_H=12.9^{+8.5}_{-7.0} \times 10^{22}~ \rm cm^{-2}$ at
$z=0.82$. In Figure \ref{optical}, plotting the (not calibrated in flux) 
2QZ optical spectrum of this
source, there is evidence for an absorption line at  $\rm
\sim4800\,\AA$. Assuming this feature is intrinsic to the QSO this is
then consistent with Mg\,{\sc ii} absorption at an outflow velocity of 
$\sim$30\,000\,km/s. This absorption line, if real, may be directly
associated with the material responsible for the X-ray absorption
(e.g. Brandt et al. 2000). We note that the strong  Mg\,{\sc ii}   
absorption feature of source \#38 may indicate a low-ionization BAL
QSO. This class of sources is believed to have high column densities
associated with Compton-thick AGNs (Gallagher et al. 2002).

Furthermore, within our sample there are another nine sources
which although may not show significant amount of absorption above the
90 per cent confidence level, their best fit value of $\Gamma$ is
quite hard, $\Gamma<1.6$ (but see note in section 4.2). The average spectrum
of these sources is estimated $\Gamma=1.5^{+0.20}_{-0.20}$,
significantly flatter than $\Gamma=1.9$  suggesting absorption. The
above photon index corresponds to an observed  column density of
$\sim10^{21}~ \rm cm^{-2}$. Given that the redshift distribution of
these sources is  between 0.5 and 2.5 the above rest frame column
density, if real, could be higher.  These sources comprise about 10
per cent of our 2QZ sample.   

It might be possible that the above 2QZ sources are associated 
with radio loud QSOs. 
Indeed radio loud QSOs  tend to have either 
intrinsically flatter photon indices 
or enhanced photoelectric column densities
(Elvis et al. 1993; Reeves \& Turner 2000).
As already discussed in
section 2, the sources above do not have radio counterparts to
the limits of the FIRST or the NVSS surveys. However for the optical
magnitude  $\rm b_{j}>18.25$ (i.e. 2QZ selection)  the radio
loud/quiet boundary lies below the  flux density limits of both the
FIRST and the NVSS surveys  (Brinkmann et al. 2000). It is therefore
possible that some of the X-ray detected sources that show evidence
for absorption are associated with radio loud AGN with radio flux
densities below the FIRST and the NVSS limits. 

We also estimate the average photon index of the 66 2QZ  QSOs 
with X-ray counterparts and find $\Gamma=2.05^{+0.07}_{-0.06}$.
There is no evidence for evolution of the photon index either with
redshift  (see Figure \ref{gamma_evol}) or flux (see section 4.3). 
With the exception of the sources discussed above this suggests that,
on average, the X-ray absorption is not important in the sample of
X-ray detected QSOs. 

In addition to X-ray spectral fitting we further explore the
X-ray absorption within our sample using the two point optical/X-ray
spectral index, $\alpha_{ox}$. Following Green et al. (1995), we
define  $\alpha_{ox} = 0.383 \log (f_{2500}/ f_{2\,\rm
keV})$, where $f_{2500}$ and $f_{2\,\rm keV}$ are the rest frame flux
densities at 2500\,\AA\ and 2\,keV respectively. We calculate $f_{\rm 
2500}$ from the $U$-band magnitudes assuming a power-law optical
spectral energy distribution of the form $f_\nu\propto
\nu^{-0.5}$. The $f_{2\,\rm keV}$ is estimated from the 0.5-8\,keV
flux assuming a photon index $\Gamma=1.7$. 
Adopting for the flux estimation \g=2 results in a negligible change
in $a_{ox}$ of $\sim 0.02$ for a source at z=2.
A high value of
$\alpha_{ox}$ suggests a low X-ray flux relative to the optical
flux. As discussed by Brandt et al. (2000) the $\alpha_{ox}$ can
reveal soft--Xray--weak candidates that are believed to be associated
with enhanced photoelectric absorption.   
In Figure \ref{aox} we plot the $\alpha_{ox}$ index versus 2500\,\AA\ 
luminosity density ($L_{2500}$). Both X-ray detected QSOs and 2QZ 
sources without
X-ray counterparts are plotted. For the latter the 3$\sigma$ 
upper limits in the X-ray flux are used. On the whole, the data do
not suggest the presence of an X-ray weak and optically bright QSO
population. The range of $\alpha_{ox}$ values  in Figure \ref{aox} is
consistent with that expected for optically selected QSOs (Green et
al. 1995). 

Most previous studies suggest that the $a_{ox}$ 
depends on 2500 {\AA} luminosity 
(see Vignali et al.  2003a and references therein).
We further explore this issue
using our homogeneously selected QSO sample. 
To take into account the lower limits in $a_{ox}$ 
we employ the {\sl ASURV} software package Rev 1.2 
(LaValley, Isobe, \& Feigelson 1992), 
which implements the survival analysis methods 
presented in Feigelson \& Nelson (1985)
and Isobe, Feigelson \& Nelson (1986). 
We use the Spearman rank order correlation 
test and the EM (Estimate and Maximize) 
regression algorithm. These statistical tests show 
that there is a strong correlation between 
$a_{ox}$ and $L_{2500}$ at the 4$\sigma$ confidence level.
The slope of the best linear fit is $0.132 \pm 0.03$ 
and the constant equals $-2.46\pm 0.93$.
These best fit parameters are in agreement with those of 
Vignali et al. (2003a) within the 1$\sigma$
standard deviation error.
In Figure \ref{aox} we plot our best linear fit 
(solid line) and that derived by Vignali et al. 2003a
(dashed line).

Only one source (\#38 in Table 1) has $\alpha_{ox}\approx2$
and therefore could be significantly X-ray obscured. 
Unfortunately, the poor photon statistics (38 counts)
do not allow us to firmly establish whether 
this source is obscured.
The  optical spectrum (not calibrated in flux)
 of this source however, presented in
Figure \ref{optical1}, 
shows a broad absorption feature at 
$\sim$6200 \AA.
 This can be interpreted  as the Mg\,{\sc ii}
absorption line at an outflow velocity of $\sim$30\,000\,km/s that may
be directly associated with any X-ray absorbing material. Brandt et  
al. (2000) found a strong correlation between the equivalent width of
the  {\sc C\,iv} absorption line and the X-ray weakness measured by
the $\alpha_{ox}$ index. They argue that the X-ray weakness in their 
sample is due to photoelectric X-ray obscuration with the absorbing
medium also responsible for the absorption features in the
ultraviolet such as the {\sc C\,iv}. Unfortunately,  the 2QZ spectral
window does not include the {\sc C\,iv} line to directly compare our
source with the Brandt et al. (2000) results. It is surprising
however, that we do not find strong evidence for X-ray  absorption in
this source.   

From the sources that show evidence for X-ray obscuration or flat
X-ray spectra only 3 (\#22, 38, 60) have $\alpha_{ox}>1.7$ suggesting
enhanced (but not extreme) absorption (Brandt et al. 2000; Gallagher
et al. 2001). The evidence above indicates that the $\alpha_{ox}$
index does not reveal {\it all} the obscured AGN candidates. This can
be attributed to: (i) X-ray flux variations affecting the estimated
$\alpha_{ox}$ index (an increase in the X-ray flux  by a factor of 3
results in a reduction of $\alpha_{ox}$ by $\sim0.18$), (ii) distinct
X-ray and UV/optical absorber (Gallagher et al. 2004), (iii) the
relative insensitivity of 2\,keV photons to column densities  lower
than $10^{22}~\rm cm^{-2}$.  In the case of our  high redshift
($z>0.5$) QSOs the latter is  a particularly plausible scenario even
for larger column densities due to the k-correction (see section 4.1).   

In addition to X-ray detections, 30 sources in our sample (about 30 per 
cent of 2QZ QSOs) remain undetected in our X-ray survey and are plotted
as lower limits in  Figure \ref{aox}. Some of these sources remain
undetected because either they are at high redshifts and hence, too 
X-ray faint for the survey limit or lie at large off-axis angles from
an XMM-{\it Newton} pointing centre where the sensitivity is
reduced. Indeed, the fraction of undetected sources drops to 17 per
cent for 2QZ QSOs at $z<2$ and off-axis angles
$<12$\,arcmin. Nevertheless, the X-ray undetected 2QZ population may
also comprise X-ray absorbed systems. We attempt to constrain the
X-ray spectral properties of these sources using stacking analysis to
estimate a mean hardness ratio of $-0.59\pm0.11$ (using MOS data)
corresponding to a low mean observed column density of about
$5\times10^{20}\,\rm cm^{-2}$. Although this result suggests that most
of the X-ray undetected QSOs are not significantly obscured we cannot
exclude the possibility that some of them have enhanced photoelectric
absorption that is not revealed in the stacked spectrum. Indeed, among
the X-ray undetected 2QZ QSOs there are two sources classified as
Broad Absorption Line (BAL) QSOs (see Figure \ref{aox}). This class of
QSOs  are weak in both the soft and the hard X-ray bands with respect
to their optical light (Brandt et al. 2000; Gallagher et al. 2001;
Green et al. 2001). This is  suggested to be due to a large column of
absorbing gas ($\sim 10^{23}~\,  \rm cm^{-2}$) rather than their
intrinsic SED. The two BAL QSOs in our sample remain undetected in our
X-ray survey, while the $3\sigma$ upper limits in their flux do not
provide strong constraints in their $\alpha_{ox}$ index.

It is worth noting that the fraction of BAL QSOs in our sample 
(both X-ray detected and undetected sources) is very low 
(about 2 per cent). 
A similarly low fraction of BAL systems is found in
the full 2QZ catalogue.
Hewett \& Foltz (2003) showed that the fraction of
BAL QSOs is doubled when considering 
selection effects such as the optical flux lost 
due to absorption. Taking these effects into account 
they found that BAL QSOs represent about 20 per cent 
of optically selected QSOs. Also BAL QSOs are, on average,
redder than UVX selected QSOs and therefore easily missed by UVX
studies (Reichard et al. 2003). The evidence above suggests that  the
2QZ survey might be biased against finding BAL QSOs. 
We also note that the fraction of BAL QSOs in the 2QZ survey 
increases (4 per cent), 
when considering only the high redshift QSOs
since the  {\sc C\,iv} absorption
line (primarily used to identify BAL QSOs) is 
entering the 2QZ spectral window only at high redshifts, $z>2$.
It is possible that a fraction of the 2QZ
QSOs in the present sample (most likely those with evidence for
X-ray obscuration or without  X-ray counterparts) may be associated
with BAL systems.

Our analysis suggests that about 10\% of the X-ray
detected QSOs (10 out of 96) show evidence for X-ray absorption. 
This is a lower limit to the fraction of optically selected QSOs with X-ray
absorption since some of the X-ray undetected 2QZ sources are likely
to be absorbed. This number however, is expected to be small as
suggested by the soft stacked X-ray spectrum of the 30 X-ray
undetected systems. The above result is consistent with that derived by 
Brandt et al. (2000). 
They  found the percentage of the low-redshift
(z$<$0.5) soft X-ray weak
QSOs in the opticaly selected QSOs population to be 11\%.
Here we have extended this result to significantly
higher redshifts.
We note that this fraction does not include all
the obscured broad line QSOs.  For example  recent studies suggest 
the presence of a reddened  X-ray absorbed broad line QSO population
(Wilkes et al. 2002; Risaliti et al. 2001). These objects have red
optical colours and are therefore missed from optically selected
samples like the 2QZ.   

\section{CONCLUSIONS}
We explore the X-ray properties of 2QZ QSOs using 
data from a wide field ($\rm \approx 2.5\,deg^2$), shallow  
[$f(0.5-8\,\rm keV)\approx 10^{-14}$\,\funits]  
XMM-{\it Newton} survey.  

The average photon index of the 66 X-ray detected QSOs is
$\sim2$ suggesting  that, on average, absorption effects are not
important in this population. On the basis of individual X-ray
spectral fittings there is evidence for significant obscuration in
only one X-ray detected optically selected QSO. This source shows an
intrinsic rest frame column density of $\rm N_H=12.9^{+8.5}_{-7.0}
\times 10^{22}\, \rm cm^{-2}$.  Additionally, a small number
(total of 9) of optically selected QSOs have X-ray spectral properties
suggesting moderate absorption ($\rm
N_H\approx10^{21}\,cm^{-2}$). This population may be associated with   
radio loud and/or BAL QSOs. Our analysis suggests a lower limit to the
fraction of optically selected QSO with X-ray absorption of about 10\%
(10 out of 96). 

However, about 30 per cent of the 2QZ QSOs remain undetected in our
X-ray survey. Although the mean X-ray spectral properties (using
stacking analysis) are consistent with a Galactic value of the 
average column density, we cannot exclude the possibility that some 
of these sources are absorbed. 

\section{Acknowledgments}
 We thank the anonymous referee for valuable comments and 
 suggestions. This work is funded by the European Union and the Greek
 Ministry of Development in the framework of the Programme
 'Competitiveness - Promotion of Excellence in Technological Development
 and Research - Action 3.3.1', Project 'X-ray Astrophysics with ESA's
 mission XMM', MIS-64564. 
 The 2dF QSO Redshift Survey (2QZ) was
 compiled by the 2QZ survey team from observations made with the
 2-degree Field on the Anglo-Australian Telescope.

\end{document}